\documentclass{pasj00}
\SetRunningHead{A. Arai}{Anti-Correlation of NIR and X-ray in GRS~1915+105}
\Received{}%{yyyy/mm/dd}
\Accepted{}%{yyyy/mm/dd}
\begin{document}
\title{
Anti-Correlation of the Near-Infrared and X-Ray Variations of 
 the Microquasar GRS~1915$+$105 in Soft State
}

\author{
Akira \textsc{Arai}\altaffilmark{1},
Makoto \textsc{Uemura}\altaffilmark{2},
Mahito \textsc{Sasada}\altaffilmark{1},
Sergei A. \textsc{Trushkin}\altaffilmark{3},
Yoshihiro \textsc{Ueda}\altaffilmark{4},\\
Hiromitsu \textsc{Takahashi}\altaffilmark{2},
Koji S. \textsc{Kawabata}\altaffilmark{1},
Masayuki \textsc{Yamanaka}\altaffilmark{2},
Osamu \textsc{Nagae}\altaffilmark{2},\\ 
Yuki \textsc{Ikejiri}\altaffilmark{1},
Kiyoshi \textsc{Sakimoto} \altaffilmark{1},
Risako \textsc{Matsui} \altaffilmark{1},
Takashi \textsc{Ohsugi}\altaffilmark{1,2},\\
Takuya \textsc{Yamashita}\altaffilmark{2}, 
Mizuki \textsc{Isogai}\altaffilmark{5},
Yasushi \textsc{Fukazawa}\altaffilmark{1},
Tsunefumi \textsc{Mizuno}\altaffilmark{1},\\  
Hideaki \textsc{Katagiri}\altaffilmark{1},
Kiichi \textsc{Okita}\altaffilmark{6}, 
Michitoshi \textsc{Yoshida}\altaffilmark{6},
Kenshi \textsc{Yanagisawa}\altaffilmark{6},\\
Shuji \textsc{Sato}\altaffilmark{7},
Masaru \textsc{Kino}\altaffilmark{7} 
and Kozo \textsc{Sadakane}\altaffilmark{8}
}

\altaffiltext{1}{
Department of Physical Science, Hiroshima University, 
Kagamiyama 1-3-1, Higashi-Hiroshima 739-8526}
\email{arai-akira@hiroshima-u.ac.jp}
\altaffiltext{2}{
Hiroshima Astrophysical Science Center, Hiroshima University, Kagamiyama
1-3-1, Higashi-Hiroshima 739-8526}
\altaffiltext{3}{
Special Astrophysical Observatory of Russian Academy of Science
(SAO-RAS), Nizhnij Arkhyz, Karachai-Cherkessia, 369167 Russia}
\altaffiltext{4}{
Department of Astronomy, Kyoto University, Sakyo-ku, Kyoto 606-8502}
\altaffiltext{5}{
Ishigakijima Astronomical Observatory, 
National Astronomical Observatory of Japapan, 
Arakawa 1024-1, Ishigaki, Okinawa 907-0024}                          
\altaffiltext{6}{
Okayama Astrophysical Observatory, 
National Astronomical Observatory of Japan, Kamogata, Okayama 719-0232}
\altaffiltext{7}{
Department of Physics, Nagoya University, 
Furo-cho, Chikusa-ku, Nagoya 464-8602}
\altaffiltext{8}{
Astronomical Institute, Osaka Kyoiku University,
Asahigaoka, Kashiwara, Osaka 582-8582}


\maketitle

\begin{abstract}
  We report detailed, long term near-infrared (NIR) light curves of
 GRS~1915$+$105 in 2007--2008, covering its long ``soft state'' for
 the first time. From our NIR monitoring and the X-ray data of the All
 Sky Monitor (ASM) onboard {\it Rossi X-ray Timing Explorer }({\it RXTE}),
 we discovered that the NIR flux dropped by $>$1 mag during
 short X-ray flares with a time-scale of days. With the termination of
 the soft state, the $H-K_{\rm s}$ color reddened and the
 anti-correlation pattern was broken.   The observed $H-K_{\rm s}$ color
 variation suggests that the dominant NIR source was an accretion disk
 during the soft state.  The short X-ray flares during the soft state
 were associated with spectral hardening in X-rays and increasing radio
 emission indicating jet ejection. The temporal NIR fading during the
 X-ray flares, hence, implies a sudden decrease of the contribution of
 the accretion disk when the jet is ejected.
\end{abstract}

\section{Introduction}
The black hole X-ray binary GRS~1915+105 is well-known and one of the
most intriguing and variable galactic sources in the X-ray, infrared and
radio bands. The object was first observed in outburst in 1992 by the WATCH
instrument on-board the {\it GRANAT} satellite 
\citep{castro-tirado92,castro-tirado94}. 
Radio observations led to the identification of a superluminal radio
source with ejecting plasma clouds at $v \sim 0.92$c
(\cite{mirabel94nat}). 
These radio jets and their superluminal motion are   
reminiscent of quasars, and hence this source is called a
``microquasar''.  Using the motion of the relativistic jets,
\citet{fender99} estimated an upper limit of the distance to be 
$11.2 \pm 0.8$~kpc.
From NIR spectroscopic observations, the companion star is
classified as a K--M type giant. Its orbital period and the mass of the
black hole were estimated to be  $33.5 \pm 1.5$ days and 
$14 \pm 4$~M$_{\solar}$, respectively (\cite{greiner01a},b).

GRS~1915$+$105 occasionally enters ``long State A (or class~$\phi$)''
categorized in \citet{belloni00}, sometimes referred as the ``soft
state''. This state is characterized by a soft X-ray spectrum with a
low flux level in the X-ray and radio regime. The origin of the soft
X-ray emission is not clear, either an emission from the accretion
disk with a high innermost temperature ($>1$ keV) and/or a Comptonized
spectrum (e.g., \cite{done04}; \cite{ueda09}). This state has
been observed once a few years and typically continued for several
weeks (\cite{reig03}). 
Since the source spends the most time in the low/hard regime (State C in
\cite{belloni00}), the occurrence rate of the soft state is low.
Although the soft state is one of the
fundamental, and hence, important phases for understanding the nature
of GRS~1915$+$105, multi-wavelength studies have poorly been performed
during this state due to its low incidence.

In August 2007, GRS~1915$+$105 newly entered the soft state after the
last one in August 2005. Here, we report our results of long and
continuous NIR photometric monitoring of GRS~1915$+$105 at
Higashi-Hirosima Observatory covering this state. Based on our data,
we reveal the NIR activity in the soft state in detail for the first
time. In \S~2, we describe our NIR observations and archived X-ray
data. In \S~3, we report new observational aspects obtained from our
monitoring; the anti-correlation of NIR and X-ray bands in the soft
state.  In \S~4, we discuss the origin of the NIR emission. Finally,
we summarize our findings in \S~5.

\section{Observations and Data}

\subsection{NIR observations with the ``KANATA'' telescope}

We performed NIR observations using TRISPEC (Triple Range Imager and
SPECtrograph) attached to the ``KANATA'' 1.5-m telescope at
Higashi-Hiroshima Observatory. TRISPEC is a simultaneous imager and
spectrograph with polarimetry covering both optical and near-infrared
wavelengths \citep{watanabe05}. We used the imaging mode of TRISPEC
with $H$ and $K_{\rm s}$ filters. Exposure times for each frame
varied night by night between 14--28 and 7--14~s in $H$- and $K_{\rm
s}$-band images, respectively, depending on the sky condition.  
We adopted a dithering technique, consisting of five sequential
exposures at different neighboring positions around the objects.
In a night, we typically took a few sequences, that is, 10--20 images in
total.

After making sky-subtracted and flat-fielded images, we measured $H$
and $K_{\rm s}$ magnitudes of GRS~1915+105 using a comparison star in
the same frame, which is located at R.A.=\timeform{19h15h09s.13},
Dec.=\timeform{+10D57'11''.1} ($H = 10.820$, $K_{\rm s} = 10.130$).
We quote the $H$ and $K_{\rm s}$ magnitudes of the comparison star
from the 2MASS catalog \citep{skrutskie06}.  We checked the constancy
of the comparison star using neighbor stars and found that they
exhibited no significant variations with amplitudes larger than 0.1
mag in $H$ and $K_{\rm s}$ band. In this paper, no correction was
performed for the interstellar extinction.

\subsection{Radio observations}
All the radio data were obtained with North sector RATAN-600 radio 
telescope attached with a continuum radiometers at 4.8 and 11.2 GHz. 

\subsection{X-ray data}
To estimate the daily flux level and hardness ratio in soft X-ray
range, we used the public ASM/{\it RXTE} (1.5--12 keV) daily-averaged
intensities, which were obtained from the MIT ASM web page
\footnote{$\langle$http://xte.mit.edu/ASM\_lc.html$\rangle$.}.
In addition, we also use BAT/{\it Swift} (15--50 keV) light curve, which
were obtained from the BAT/{\it Swift} team web page 
\footnote{$\langle$http://swift.gsfc.nasa.gov/docs/
swift/results/transients/\\index.html
$\rangle$.}.

\section{Results}
Our observations are shown in figure~1.  The panel~(a) of figure 1
shows the $K_{\rm s}$-band light curve obtained with TRISPEC/KANATA.
The panel~(b) depicts the temporal variations of $H-K_{\rm s}$ colors.
In the panel~(c), (d) and (e), we show the X-ray data; the soft X-ray
light curve (1.5--12~keV) and the hardness ratio (``HR2'' defined as
the flux ratio of $f_{\rm 5-12~keV}/f_{\rm 3-5~keV}$) obtained by
ASM/{\it RXTE}, and the hard X-ray light curve (15--50~keV) by
BAT/{\it Swift}, respectively.  We first describe the overall X-ray
behavior of the object during our NIR monitoring, and then, report
correlations of the NIR and X-ray fluxes.

\subsection{X-Ray behavior during our NIR monitoring}
Between  MJD~54250 and 54320, GRS~1915$+$105 was in an active state with
hard X-ray spectrum. A large flare occurred between MJD~54280 and 54320,
accompanied with a large HR2 of $\sim 1.5$.

GRS~1915$+$105 entered the ``soft state'' (state A) in $\sim$
MJD~54321, as indicated by the small hardness ratio and the low level
flux of the soft and hard X-rays.  According to \citet{belloni00}, the
soft state is defined by HR2$\lesssim 1.1$.  With this definition, the
object stayed in the soft state almost all times from MJD~54321 to
54571.  We note that the duration of this soft state is $\sim 250$~d,
much longer than those of the soft states previously observed ($\sim
60$~d in \cite{ueda06}).

During the soft state, the X-ray flux occasionally showed modulations 
and flares having a time-scale of days. At the peak of an X-ray flare 
on MJD~54368, the hardness ratio, HR2, was $1.3$, which is higher than
that in the soft state. Also in the other X-ray flares and modulations, 
we can see that HR2 was high in a range of $1.2$ -- $1.5$.  
In the flare on MJD~54368, the flux increased by a factor of 3.4 and 
17.7 in the ASM and BAT count rate, respectively.  While the amplitudes
in the flux density depend on the spectral index during the flare, the
large HR2 and the large amplitude in the BAT count rate indicate that
the X-ray flares had X-ray spectra harder than that in the ordinary soft
state.  Figure~\ref{lc2} shows the light curves in the X-ray, radio, and NIR 
ranges between MJD~54320 and 54450.  On MJD~54368, at the peak of the 
X-ray flare, the radio flux significantly increased from $\sim 3$ mJy
to $223$ mJy at 4.8~GHz and from $\sim$ 17 mJy to 202 mJy at 11.2~GHz.

In previous radio observations of GRS~1915+105, three types of radio
flares are known (\cite{eikenberry00}).
The class A flare is characterized by long durations (several days)
and large amplitudes. It is believed to be associated
with major jet ejections (\cite{mirabel94nat}; \cite{fender99}). 
The class B and C flares are consecutive short  ($\sim 500$---$1800$~s) flares 
(e.g. \cite{fender97}; \cite{mirabel98}; \cite{feroci99}; \cite{eikenberry00}).
Thus, the flare on MJD~54368 was presumably a class A flare, associated
with the jet ejection, based on the duration and amplitudes of the flare.

The soft state was terminated by the flare on $\sim$MJD~54571.  After this 
flare, the hardness ratio abruptly increased on $\sim$MJD~54589, which 
indicates the occurrence of a state transition.  The increase of the hard 
X-rays (the panel~(e) of figure~1) also supports that the X-ray spectrum 
changed from the soft to the hard one.  This state can be categorized 
as the ``plateau state'', which is characterized by the prolonged hard X-ray
emission associated with the steady synchrotron radio emission (e.g.,
\cite{foster96}). Plateau states have a precursor flare (e.g.,
\cite{Klein-wolt02}). The plateau state in 2008 was also associated
with a pre-plateau flare \citep{trushkin08}.

Except for the atypically long duration of the soft state, the overall
X-ray behavior of GRS~1915$+$105 was a standard one in terms of the 
characteristics of the light curve (\cite{belloni00}).  Owing to the
long duration, we can investigate the correlation of the X-ray and the
NIR fluxes during the soft state for the first time in detail.  

 \begin{figure}
   \begin{center}
     \FigureFile(85mm,85mm){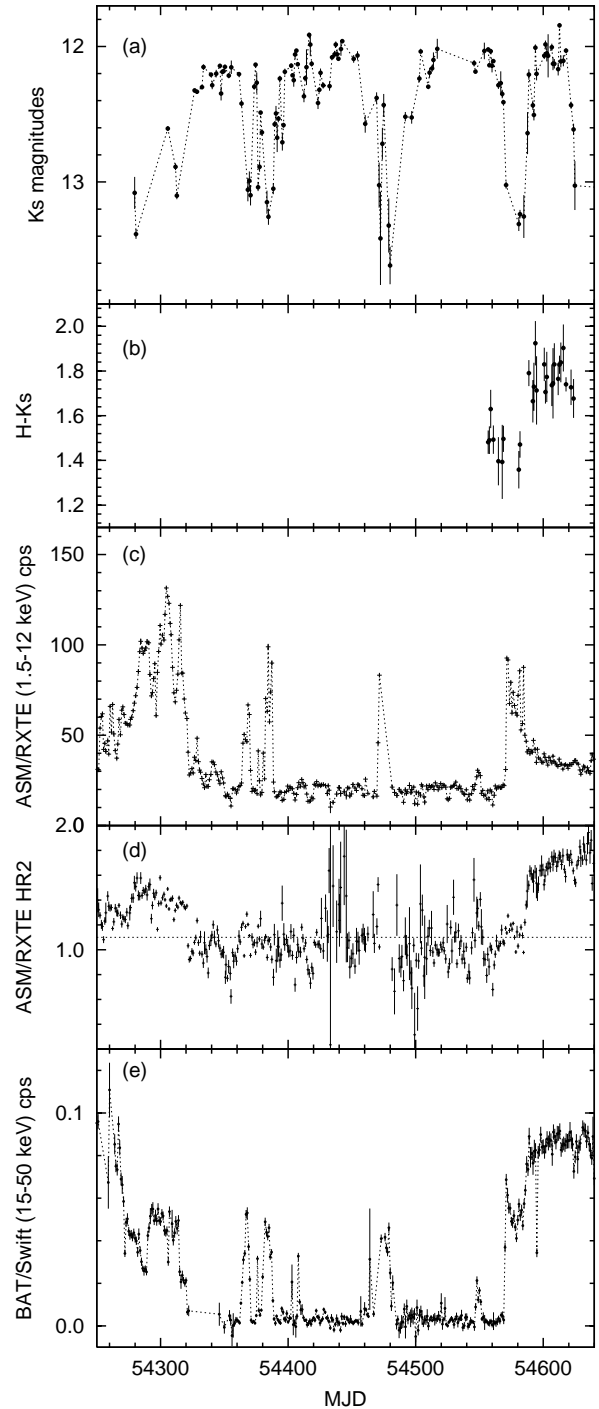}
       \end{center}
   \caption{Light curves and the temporal variations of the NIR color and 
 X-ray hardness ratio of GRS~1915$+$105.  The abscissa denotes the time 
 in MJD.  (a) $K_{\rm s}$-band light curve.   
 (b) $H-K_{\rm s}$ color variations.   The reddening correction was not
  performed.  (c) Soft X-ray light curve observed by  ASM/{\it RXTE}.
  (d) Hardness ratio (HR2) by ASM/{\it RXTE}. According to
  \citet{belloni00}, the object is considered to be in the soft
  state if HR2 is lower than the horizontal dashed line. 
 (e) Hard X-ray light curve by BAT/{\it Swift}.

}\label{lc}
\end{figure}

\subsection{Anti-correlation of the X-ray and NIR flux during the soft state}

Before the soft state, the $K_{\rm s}$-band flux was variable,
possibly correlating with the X-ray flare on MJD~54280--54320, while
our $K_{\rm s}$-band observation was too sparse to be conclusive for
the possible positive-correlation. The $K_{\rm s}$-band flux, then,
increased to $K_{\rm s} \sim 12.2$ when the object entered the soft
state and the X-ray flux reached a low level on $\sim$ MJD~54321.

In the soft state, we discovered a clear anti-correlation between the
$K_{\rm s}$-band and X-ray fluxes. In figure~1, the anti-correlation
is evident during the prominent X-ray flares in MJD~54363--54370,
54382--54387, and 54470--54480, in which the $K_{\rm s}$-band flux
decreased when the X-ray flare occurred. Furthermore, we confirmed
that the anti-correlation was present even for lower amplitude X-ray
modulations, as indicated by the dashed lines in figure~2.
A simple calculation of a correlation function suggests no significant
time-lag between the X-ray and $K_{\rm s}$-band flux over a day.
There might be a possible time-lag shorter than a day, which is,
however, difficult to be established with our available data.

This is the first time that such a clear anti-correlation was observed
between the X-ray and NIR flux on a short time-scale of days during
the soft state. We found that the similar anti-correlation behavior was
probably seen in the light curve reported in \citet{neil07}, which
observed the soft state in 2005 August. This anti-correlation behavior
is, hence, probably a common feature for the soft state in GRS~1915$+$105.

On MJD~54570, the $K_{\rm s}$-band flux decreased associated with the
X-ray flare just before the state transition from the soft state to
the plateau state. The object, then, re-brightened to 
$K_{\rm s}=12.1 \pm 0.1$ and stayed in that level during the plateau state. 
As can be seen from the panels~(a) and (e) of figure 1, the
anti-correlation feature between the NIR and the X-ray flux was
apparently broken after this state transition; the X-ray flux in the 
plateau state was higher than that in the soft state, while the NIR
flux in the plateau state kept a high level as in the soft state.

As shown in the panel (b) of figure~\ref{lc}, the $H-K_{\rm s}$ color
was relatively blue during the soft state ($H-K_{\rm s}=1.47 \pm
0.08$).  The color abruptly changed to be redder on MJD~54584 when the
X-ray state transition started, as shown in the hardness ratio
variation (the panel~(d) in figure~1).  During the plateau state, the
average of the color was $H-K_{\rm s}=1.77 \pm 0.07$, significantly
redder than that in the soft state.

These results indicate that the dominant NIR emission source changed due to 
the state transition. In other words, the state transition occurred 
not only in the X-ray regime, but also in the NIR one.  It is 
interesting to note that the maximum $K_{\rm s}$-band flux level in the 
plateau state was quite similar to that in the soft state, both at 
$K_{\rm s} \sim 12.1$, although the dominant emission source changed.

 \begin{figure}
  \begin{center}
   \FigureFile(85mm,80mm){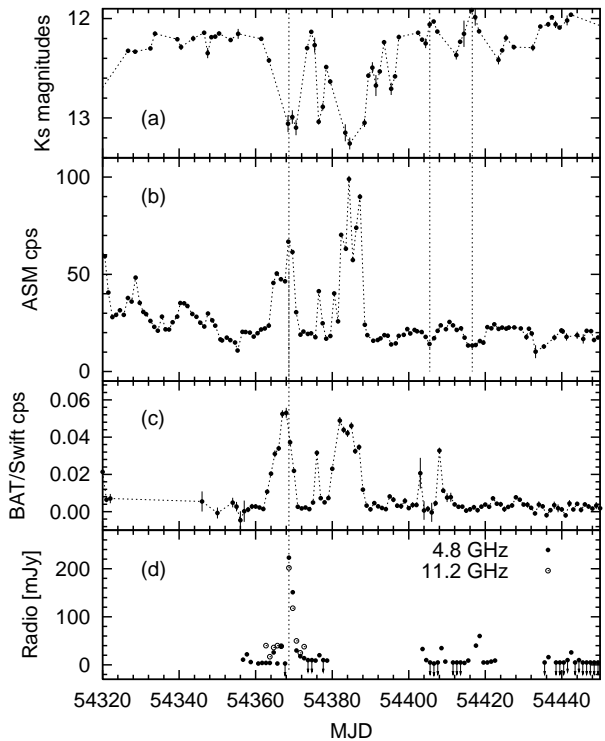}
  \end{center}
  \caption{Multiwavelength light curves between MJD~54320 and
  54450. (a) $K_{\rm s}$-band light curve. 
  (b) Soft X-ray light curve observed by ASM/{\it RXTE}. 
  (c) Hard X-ray light curve observed by BAT/{\it Swift}.  
  (d) Radio light curves in 4.8~GHz and 11.2~GHz. 
    In the panel (d), the arrows of 4.8~GHz observation indicate the
  upper limit. The vertical dashed lines are shown in order to readily
  see the anti-correlation of the NIR and X-ray fluxes.
  }\label{lc2}
 \end{figure}

\section{Discussion}

In ordinary X-ray binaries, the NIR sources can be companion stars,
accretion disks, or jets (e.g; \cite{haswell01}; \cite{chaty03}).
Here, we discuss which component was dominant and responsible for the
intriguing anti-correlation of the NIR and X-ray variations in the
soft state.

The most promising candidate is the accretion disk. First, the
accretion disk can be variable with a short time-scale of days.
Second, the $H-K_{\rm s}$ color was bluer in the soft state than that in
the plateau one. 
We can expect that the NIR emission originates from the outermost part of
the accretion disk ($T \sim 10^4\,{\rm K}$), whose $H-K_s$ color is
presumably bluer than those of the companion star (K--M giant) and the
synchrotron jet.

The jet is less likely for the NIR source in the soft state because
the radio emission was quite weak between 54356--54450, except for the
period of the X-ray flares on $\sim$~MJD~$54365$. During the X-ray
flare, figure~2 shows an anti-correlation of the NIR and radio flux.
The radio flare was presumably a class A flare, which indicates the
ejection of jets, as reported in \S~3.1.
If the NIR flux is dominated by the radiation from jets, the NIR and
radio fluxes should show positive correlations throughout the soft state.
Our results, therefore, have a negative implication for the jet
scenario for the NIR flux in the soft state. 
We can reasonably reject the companion star origin because
it is hard for a K--M-type giant star to be variable with the
amplitude of $\sim 1\,{\rm mag}$ within a few days.  Moreover, there
is no sign for orbital period variations in our light curve.

By contrast, after the transition to the plateau state, the dominant
NIR source could be changed from the accretion disk to the jet, as
suggested by previous authors (\cite{ueda02}; \cite{fuchs03};
\cite{eikenberry08}). In fact, the radio emission during the plateau
state is considered to originate from the steady, compact jet
(\cite{foster96}).

The temporal fading of the NIR flux during the X-ray flares,
thus, implies that the contribution of the accretion disk decreases when
the jet is ejected.  The mechanism of this anti-correlated variation is,
however, totally unknown.  If the soft state of GRS~1915$+$105 is
analogous to the ``high/soft state'' of ordinary black hole candidates,
the standard accretion disk can be the dominant source in the NIR to
X-ray band \citep{shakura_sunyaev76,esin97}. 
In this case, the NIR flux should correlate with the X-ray one because
both fluxes are a function of the mass accretion rate in the disk, as
reported in \citep{ueda02}.  This is opposite to our observations. We,
thus, need alternative factors for understanding the anti-correlation
feature, for example, the irradiation and/or reprocess by X-ray emission
at the outermost part of the accretion disk, or the occultation of the
NIR source by the jet or disk. 
If the irradiation or reprocess effect plays a key role, 
a short time-lag may be observed between the NIR and 
X-ray variations because of long heating and cooling 
time-scales in the outer part of the disk.
Dense multi-wavelength observations are required to reveal the 
nature of the observed anti-correlation.

\section{Summary}
We reported detailed NIR behavior of GRS~1915$+$105 during the soft
state for the first time.  The object entered the soft state 
(long State A) in August 2007.  Our results revealed a clear
anti-correlation between the NIR and X-ray fluxes in the soft state.
This feature was also confirmed for small amplitude modulations. After
the termination of the soft state, the object entered the plateau state,
where the anti-correlation pattern was broken.  
Based on the $H-K_{\rm s}$ color variation, we propose that the dominant
NIR source was an accretion disk in the soft state. Since the X-ray 
flares during the soft state were associated with the jet ejection, 
our observation suggests that the contribution of the accretion disk
decreases when the jet is ejected.
\\
\\
This work was partly supported by a Grand-in-Aid from the Ministry 
 of Education, Culture, Sports, Science, and Technology of Japan 
 (19740104,17340054).
Research has made use of BAT/{\it Swift} transient monitor results provided by
the  BAT/{\it Swift} team.

\end{document}